\documentstyle[preprint,eqsecnum,epsfig,aps]{revtex}
\tightenlines

\begin{document}
\draft
\title
{Comparison of 3-Dimensional and 1-Dimensional Schemes in the calculation of
Atmospheric Neutrinos}

\author{M. Honda, T. Kajita}
\address{Institute for Cosmic Ray Research, University of Tokyo, 
Kashiwa 277-8582,  Japan.}
\author{K. Kasahara}
\address{Dept. of Electronic Information Systems, Shibaura Inst. of Tech. 
Fukasaku, Ohmiya 330-8570,  Japan.}
\author{and, S. Midorikawa}
\address{Faculty of Engineering, Aomori University, Aomori 030-0943, Japan.}
\date{\today}

\begin{abstract}
A 3-dimensional calculation of atmospheric neutrinos flux is presented, and
the results are compared with those of a 1-dimensional one.
In this study, interaction and propagation 
of particles is treated in a 3-dimensional way including the curvature of
charged particles due to the geomagnetic field, which is
assumed to be a dipole field. The purpose of this paper is limited to 
the comparison of calculation schemes. The updated flux value with new 
interaction model and primary flux model will be reported 
in a separate paper.

Except for nearly horizontal directions, the flux is very similar to the
result of 1 dimensional calculations.
However, for near-horizontal directions an enhancement of the neutrino
flux is seen even at energies as high as 1 GeV.
The production height of neutrinos is lower than the prediction by 
1-dimensional calculation for near-horizontal directions,
and is a little higher for near-vertical directions.
However, the difference is not evident except for near-horizontal
directions.

\end{abstract}

\pacs{95.85.Ry, 14.60.Pq, 96.40.Tv}
\maketitle

\section{Introduction:}
\label{intro.sec}

The conflict between experimental data and
the theoretical predictions of atmospheric neutrinos gives evidence for
neutrino oscillations\cite{sk} 
(see also \cite{fukuda}\cite{imb}\cite{soudan2}\cite{macro}).
The data from Super-Kamiokande, which dominate the statistics 
in the atmospheric neutrino data, are well explained by 
$\nu_\tau \leftrightarrow \nu_\mu$ oscillation with 
$\Delta m^2 \simeq 3\times 10^{-3}$~eV$^2$ and $sin^2 2\theta \sim 1$.
We note that the oscillation mode of $\nu_\tau \leftrightarrow \nu_\mu$ with 
$\Delta m^2 \sim 1 \times 10^{-2}$~eV$^2$ was 
suggested\cite{lpw}\cite{bw}\cite{hhm} immediately after the discovery 
of the atmospheric neutrino anomaly\cite{hirata}, using the 
the atmospheric neutrinos flux predicted in the 1-dimensional
approximation\cite{gaisser-old}.
The theoretical study of the atmospheric neutrinos
has also been improved since that time,
but most of them still employ the 1-dimensional 
approximation\cite{gaisser-new}\cite{hkkm}.
For further study of neutrino oscillations, a better prediction of the 
atmospheric neutrino flux calculated using a 3-dimensional scheme 
may be needed.

The `3D-effects' are not so large;
the bending of muons in the geomagnetic field is $\sim 0.1$ radian 
($\sim$ 5 degree) in the average muon lifetime, and 
the transverse momentum of a secondary particle in a hadronic interaction 
is typically 0.3GeV/c.
Both are small effects for neutrinos with energy of $\gtrsim$ 1GeV,
and could be ignored for $\gg$ 1 GeV.
Therefore it is considered that 1-dimensional calculation is sufficient
for the confirmation of neutrino oscillation and their nonzero masses.
The effects become important, however, for neutrinos with energies
$\lesssim$ 1GeV.

One of the difficult problems of the 3-dimensional calculation
is the computation time.
If we sample the cosmic ray uniformly over the surface of the Earth,
roughly speaking, only $(Detector-size/Earth-radius)^2$ of the produced
neutrinos go through the detector.
The 3-dimensional calculations that have been reported so far adopt
some ideas which address this computation problem.
Tserkovnyak et al. assumed a huge detector size\cite{tserk}.
However, they still suffered from small statistics.
On the other hand, 
Battistoni et al. assumed a spherical symmetry 
ignoring the geomagnetic field in the air\cite{battis},
and they found an enhancement of atmospheric neutrino fluxes 
in near-horizontal directions at low energies.
This is a general feature in the 3-dimensional calculation, as
Lipari gives an explanation of it in terms of geometry\cite{lipari-ge}.
Such a feature is not seen in the 1-dimensional calculation.

In this study, we introduce a dipole geomagnetic field 
both for the geomagnetic cutoff test and particle simulation in the air.
With the axisymmetry of the dipole geomagnetic field,
we can integrate the results over longitudinal directions and
reduce the computation time to get the sufficient statistics.
This dipole magnetic field may be an over simplification for the 
geomagnetism, 
but is useful to estimate the effect of the geomagnetic field 
in the air on atmospheric neutrinos.
This is stressed by Lipari\cite{lipari-ew}.

In this paper, we concentrate on the comparison of the 3-dimensional 
calculation, with and without the geomagnetic field in the air, and the
1 dimensional one. The flux value will be reported in a separate paper
with an improved interaction model.

\section{Simulation Setups and procedure}
\label{setup.sec}

\begin{figure}[tbh]
\centerline{\epsfxsize=12cm\epsfbox{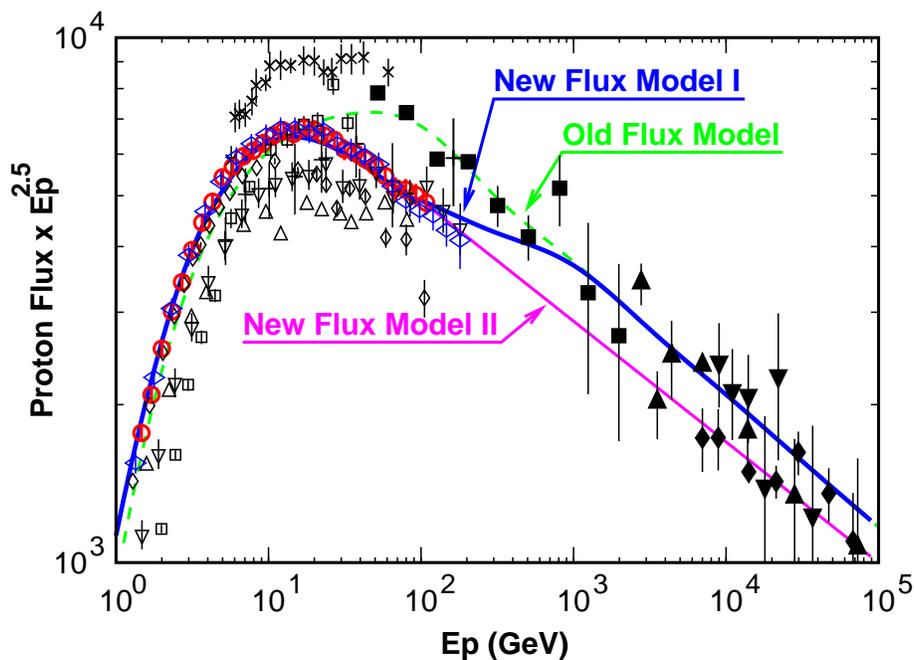}}
\caption{Primary cosmic ray observation and our model curves for protons
at solar minimum. New Flux Model I is used in this study.
Crosses indicate data from Ref.[18],
open squares MASS[19],
open upward triangles LEAP[20], 
open downward triangles IMAX[21],
open vertical diamonds CAPRICE[22], 
open circles BESS[23], and
open horizontal diamonds AMS[24].
Pluses indicate data from Ref.[25]
closed squares Ref.[26]
closed vertical diamond JACEE[27]
closed upward triangles Ref[28], and
closed downward triangles Ref[29].
}
\label{primary}
\end{figure}

%primary flux model.
The atmospheric neutrino flux of $\lesssim$~1~GeV, in which we
expect sizable differences between 1-dimensional and 3-dimensional 
calculations, is mainly produced by the cosmic rays with energies 
below 100~GeV.
The recently observed proton cosmic ray flux in this energy region
is lower than the HKKM flux model above 30 GeV, 
showing the maximum difference of 30~\%
at around 100~GeV (Fig.~\ref{primary}).
Variations of observed flux for $<$~10~GeV are also seen.
However, this is mainly due to modulation by Solar activity,
and they agree with each other when a proper correction is applied.
Therefore, we renew the primary proton flux model based on recent 
observations, especially those of BESS\cite{bess} and AMS\cite{ams}.
Above 100~GeV, we construct two primary flux models for proton cosmic 
rays: New Flux Model I and New Flux Model II shown in Fig.~\ref{primary}
with available data. 
They are considered as the upper and lower bounds
of plausible extrapolation from lower energies.
The fraction of heavier chemical composition in cosmic rays
is small at $<$ 100 GeV, and we take the values of the old HKKM flux model
for them.

As a temporary choice, we use the New Flux Model I in this study.
The differences in the primary flux model, 
including that for heavier nuclei, would not result in large
differences in the comparison of 1D and 3D calculation schemes.
However, the differences in neutrino flux between the New Flux Model I, II, 
and the Old Flux Model are briefly addressed in a later section.

%interaction model
For the interaction model, we use the same interaction model as
the HKKM\cite{hkkm} calculations in this study.
We stress that, however, we are improving the interaction model,
since the combination of new flux model and present interaction model does 
not explain the observed flux of secondary cosmic rays 
at several altitudes.
The updated interaction model as well as the primary flux model 
will be reported in a forthcoming paper with the resulting neutrino flux.

We assume that the surface of the Earth is a simple sphere with 
radius of $R_e \simeq 6378$~km, 
and use a geomagnetic coordinate system such that the center of the Earth is
the origin and the line from the center of the Earth to the magnetic 
north pole is the z-axis.
The geomagnetic field is approximated by a dipole magnetic field as,

\begin{equation}
B_x = B_0\cdot 3 zx R_e^3/r^5, {~~~~}
B_y = B_0\cdot 3 zy R_e^3/r^5, ~~~~~{\rm and~~~~}
B_z = B_0\cdot (3z^2-r^2) R_e^3/r^5
\end{equation}
with $B_0=-0.30$ Gauss in this coordinate system.
The position of magnetic north pole is calculated to be at (71.4W, 79.3N)
for the geomagnetic field in 1995.
And the magnetic latitudes for SK, Soudan-II, and SNO are
26.9~N, 58~N, and 54.0~N, respectively.
The position of SK  may be considered as the mid-magnetic-latitudes (MML),
and that of Soudan-II and SNO as the high-magnetic-latitudes (HML).

\begin{figure}[tbh]
\centerline{\epsfxsize=11cm\epsfbox{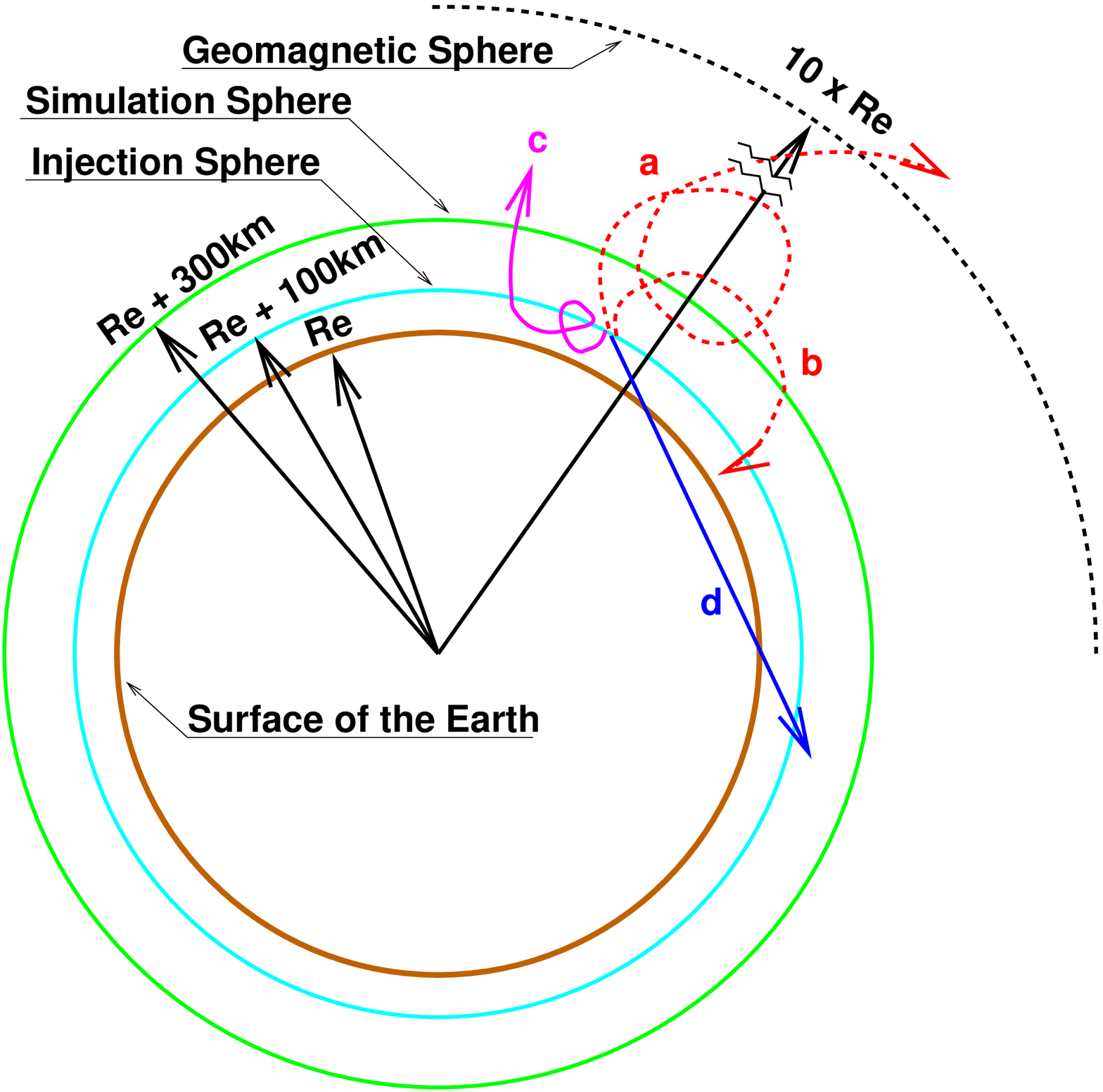}}% 
\caption{Schematic view of the 3-dimensional calculation of atmospheric 
neutrinos. The curves in the figure show  a) the backtracking orbit for 
an allowed path, b) same as a) but for a forbidden path, 
c) the orbit of a discarded particle, and d) the track for neutrinos.}
\label{schematic}
\end{figure}

In addition to the surface of the Earth, we consider three more spheres.
The first one is called the injection sphere with the radius 
of $R_e + 100$~km,
the second one is the simulation boundary sphere with
radius of $R_e + 300$~km,
and the third one is the geomagnetic sphere with radius of $10 \times R_e$.
(See Fig.~\ref{schematic}.)
The simulation of cosmic rays starts at the injection sphere,
and is carried out in the space between the surface of the Earth 
and the simulation boundary sphere.
Outside of the geomagnetic sphere,
we consider the cosmic rays are free from the effects of the 
geomagnetic field. 

First, the chemical composition and energy are sampled for a cosmic 
ray following the energy spectrum of each chemical composition.
Then the injection position is sampled uniformly over the injection sphere
and the arrival directions are sampled so that the zenith angle distribution
is proportional to [~$\cos\theta \cdot d \cos\theta$~],
where $\theta$ is the zenith angle at the injection point. 
We record the sampled energy, the chemical composition, 
the position of the injection and the direction of the cosmic ray, 
irrespective of the result of the geomagnetic cutoff test.

For the geomagnetic cutoff test,
we trace the backward path of a particle with the same mass and energy 
as the cosmic ray but with the opposite charge.
If the particle goes out of the geomagnetic sphere within 100 sec 
without going into the injection sphere,
we judge the cosmic ray has passed the geomagnetic cutoff.
The above geomagnetic cutoff test only picks up cosmic rays which 
arrive in the injection plane for the first time.
Note that the geomagnetic cutoff test works exactly the same way
for two particles with the same rigidity ($p/Z$).

The cosmic rays which pass the geomagnetic cutoff test are fed into 
Cosmos simulation code\cite{kasahara}.
When a neutrino is produced in the simulation, the production position 
and the direction are recorded. Other particles are traced
until they decay, they leave the simulation sphere or enter the Earth. 

For the neutrinos, we calculate both the point of entrance into the Earth
and the point of emergence from the Earth; neutrinos which do not enter the 
Earth are discarded.
At each point, the arrival zenith angle and the azimuth angles 
are calculated.
We note that the arrival zenith angle is defined as the angle
between the downward normal vector direction and direction of 
the neutrino at each point.
The azimuth angle is defined as the projection angle 
on the tangential plane at each point.

We refer to the calculation setup explained in the above as the 3D 
calculation. 
We performed three other calculations with different setups to the above.
The first is a 1-dimensional calculation such that the geomagnetic cutoff test 
is applied with the same dipole magnetic field as in the 3-dimensional case,
but all particles are treated by Cosmos in a 1-dimensional fashion 
(1D calculation), in which 
all secondary particles are produced in the direction of the primary
cosmic ray.
The second is such that the geomagnetic cutoff and the interaction are treated 
in the same way as the 3D calculation 
but the effect of the geomagnetic field is ignored
in the air (3D-nomag calculation).
The third is another 1D calculation such that most of procedures are 
the same as the 1D, but the the geomagnetic cutoff is 
applied with the multi-pole expanded (8th order) geomagnetic field
(1D-multipole calculation).
This is almost the same as the HKKM calculation\cite{hkkm} except for the Flux 
model.
The New Flux Model I is used for all the above calculations.
These calculations are perfomed both for the SK site and for North America.

\section{Simulation Results}
\label{result.sec}

\subsection{Direction averaged flux}
\label{alldir}

\begin{figure}[tbh]
\centerline{\epsfxsize=12cm\epsfbox{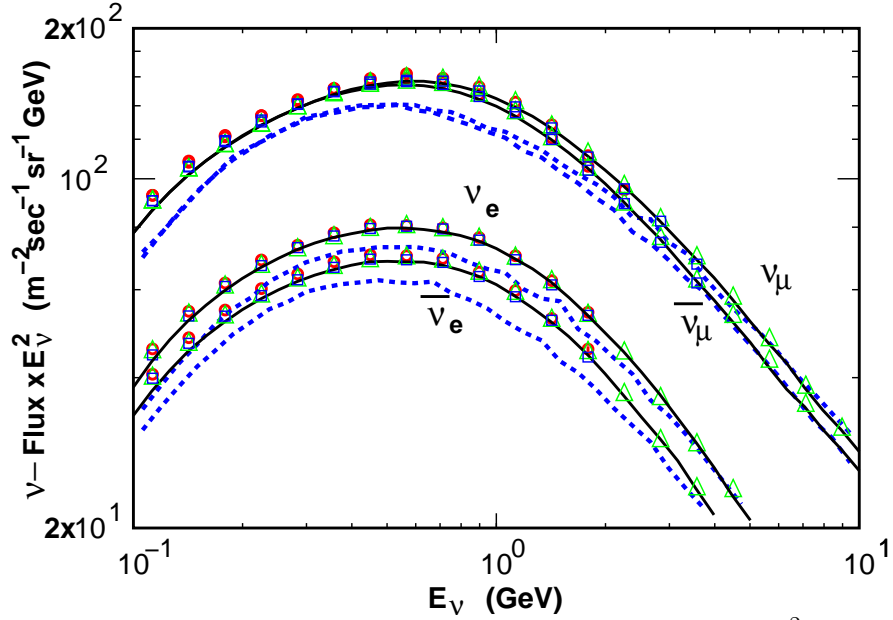}}
\caption{Direction averaged atmospheric neutrino flux multiplied by 
$E_\nu^2$ for the MML. 
The squares are for 3D, triangles for 1D, and circles for 3D-nomag
calculations. 
The solid and dashed lines show the neutrino fluxes of 1D-multipole and 
Battistoni et al.~[13] respectively.}
\label{kam-alldir}
\end{figure}

\begin{figure}[tbh]
\centerline{\epsfxsize=12cm\epsfbox{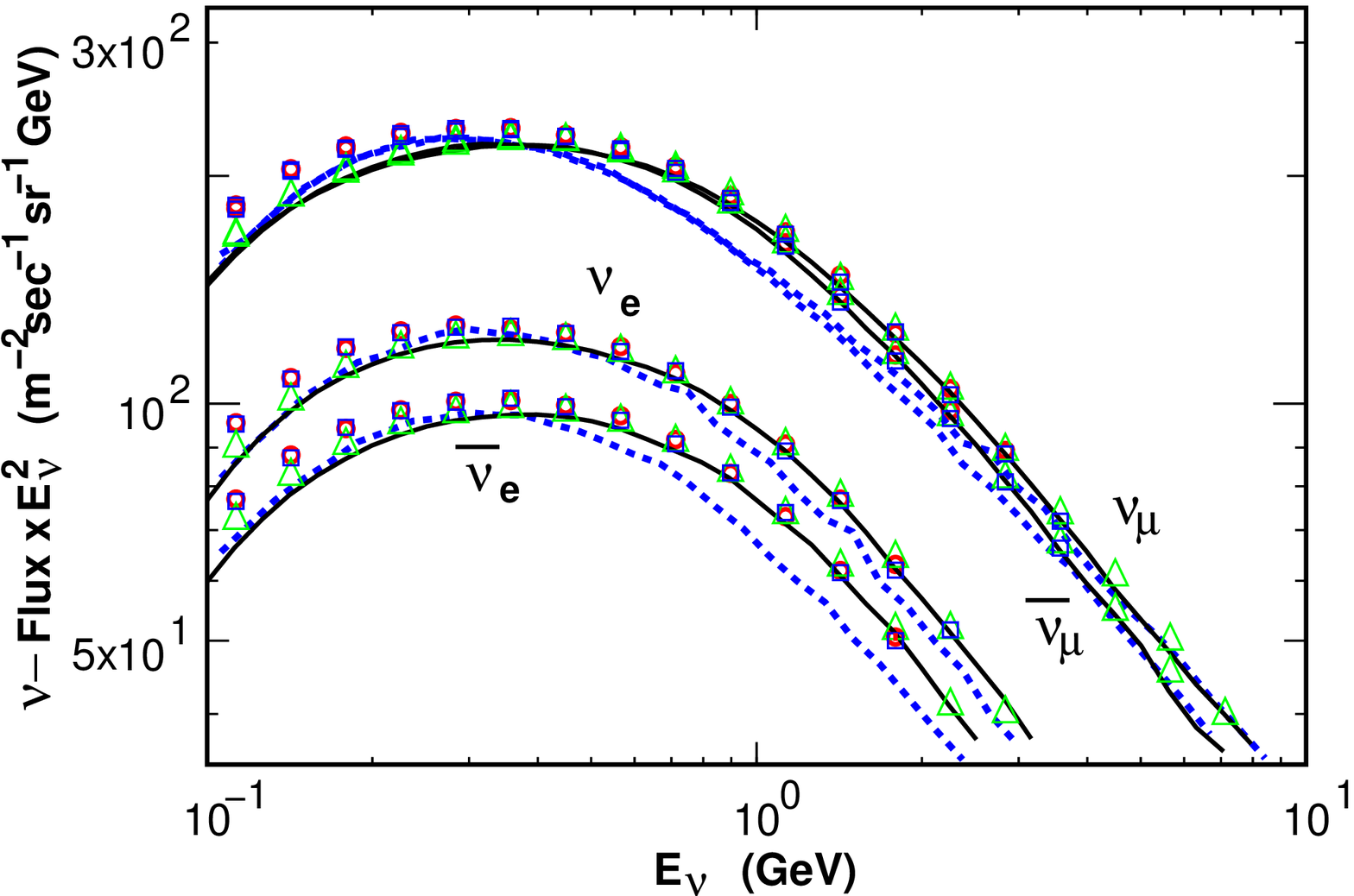}}
\caption{Direction averaged atmospheric neutrino flux multiplied by 
$E_\nu^2$ for the HML. 
The squares are for 3D, triangles for 1D, and circles for 3D-nomag
calculations. 
The solid and dashed lines show the neutrino fluxes of 1D-multipole and 
Battistoni et al.~[13] respectively.}
\label{sno-alldir}
\end{figure}

\begin{figure}[tbh]
\centerline{\epsfxsize=17cm\epsfbox{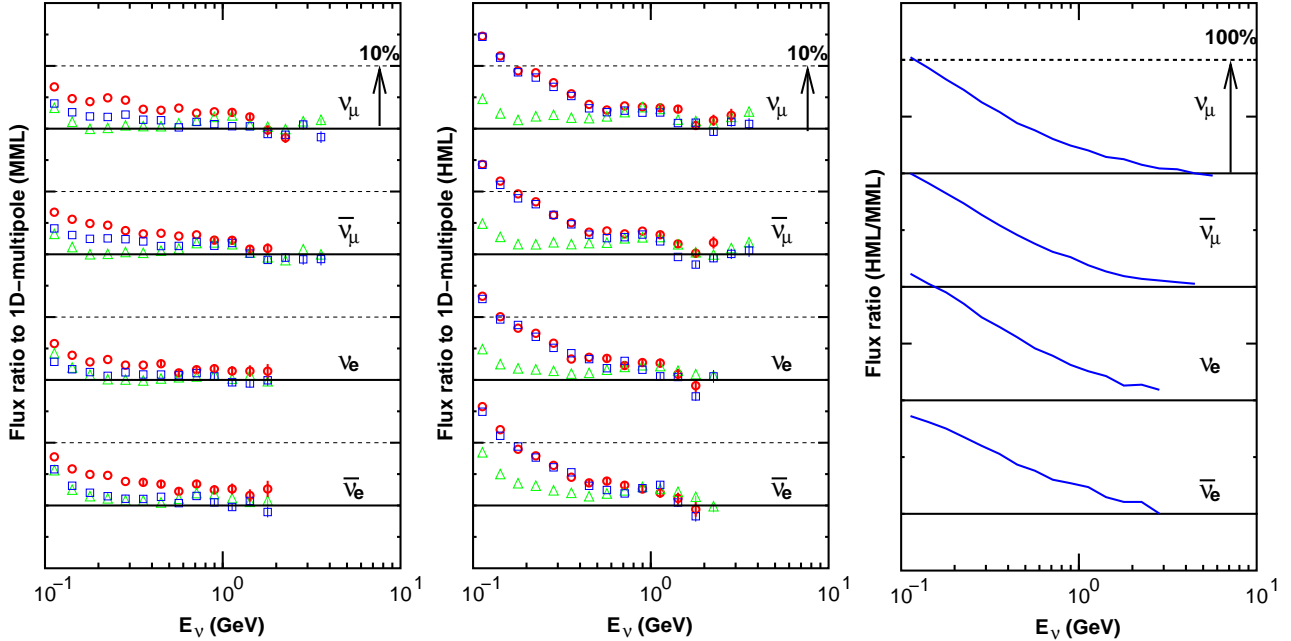}}
\caption{Flux ratios of 1D, 3D, and 3D-nomag to the 1D-multipole
in MML(left) and HML(center),
and the ratio of 3D for HML to MML (right).
The squares are for 3D, triangles for 1D, and circles 
for 3D-nomag calculations. 
}
\label{alldir-ratios}
\end{figure}

In Fig.~\ref{kam-alldir}, 
we show the energy spectra averaged over all directions 
of atmospheric neutrinos predicted by several different 
calculations for the MML (mid-magnetic-latitudes), 
and in Fig.~\ref{sno-alldir} for HML (high-magnetic-latitudes).
To see the difference more clearly, we also calculated
the ratios of the 1D, 3D, and 3D-nomag to the 1D-multipole fluxes 
for MML and HML,
and the ratio of 3D fluxes for HML to that for MML 
in Fig.~\ref{alldir-ratios}.

The differences between 1D, 3D, 3D-nomag, and 1D-multipole calculations 
using the same primary cosmic ray flux model are small for MML.
The differences are less than 5~\% for $\gtrsim$ 0.2~GeV for all 
the calculation schemes, and for $\gtrsim$ 0.1~GeV between
1D and 3D.
On the other hand,
the differences between different calculation schemes are 
larger in HML than those in MML.

The neutrino flux difference between the
New Flux Model-I and II is $\lesssim$ 2~--~3~\% at 1~GeV and it grows to 
$\sim$ 10~\% at 10~GeV.
It is rather small in the energy region where
the `3D-effects' are important.
However, the difference between the Old Flux Model and New Flux Model-I is 
8~$\sim$~12~\% at 1~GeV and grows to a maximum of $\sim$~20~\% at 
6~$\sim$~8~GeV, and decreases above these energies.

The flux of Battistoni et al.\cite{battis} is a little smaller than ours
for $\lesssim$ 3~GeV, and the difference is $\sim$ 15\% at 1~GeV in MML.
In HML, their flux is again smaller than ours 
for 0.3~GeV $\lesssim$ E$_\nu \lesssim $ 3~GeV ($\sim$ 15\% at 1~GeV).
But below 0.3~GeV, their flux is similar or even larger than ours.
However, their flux is very similar to ours
at energies $\gtrsim$ 3~GeV in both MML and HML.

\subsection{Zenith angle dependence}
\label{sec-zdep}

\begin{figure}[tbh]
\centerline{\epsfxsize=17cm\epsfbox{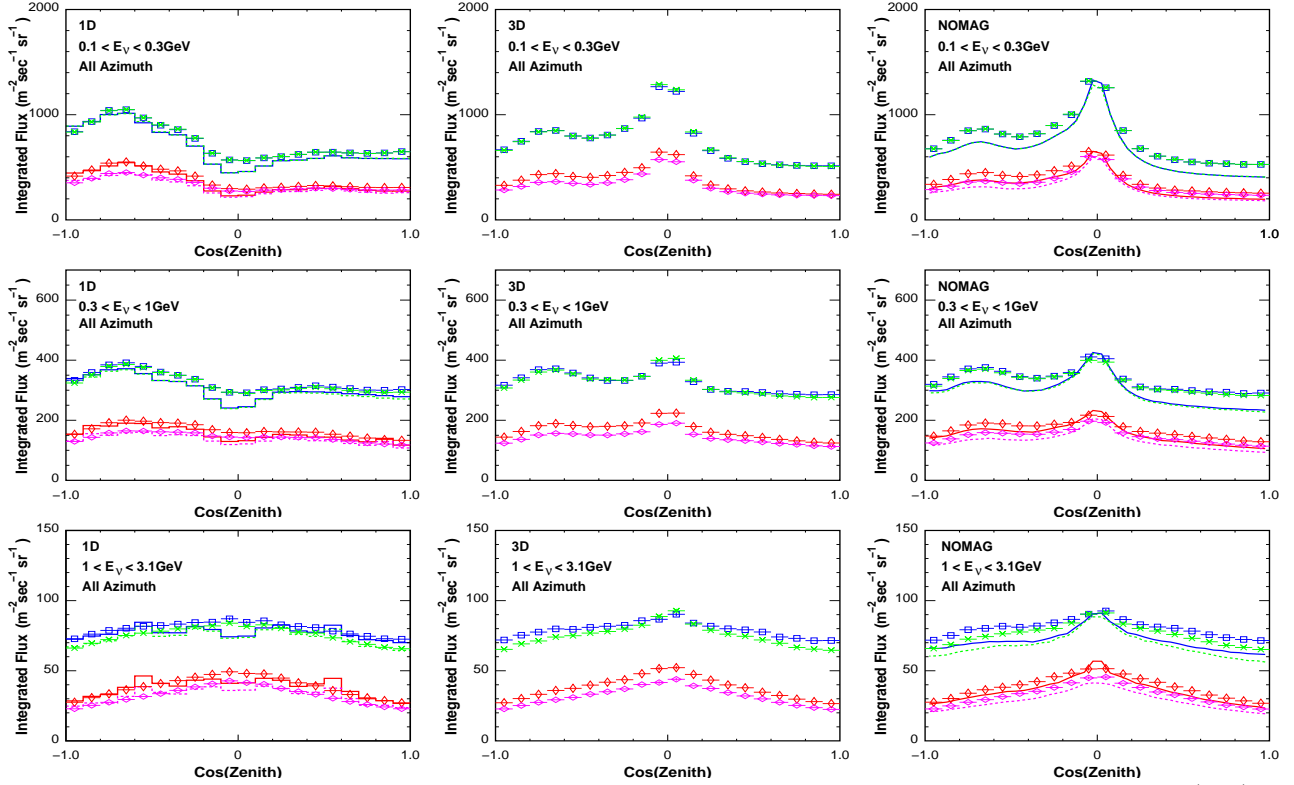}}
\caption{Zenith angle dependence of atmospheric neutrino flux
calculated in the 1D(left), 3D(center), and 3D-nomag(right) schemes for MML.
Squares are for $\nu_\mu$, asterisks for $\bar \nu_\mu$, vertical diamonds
for $\nu_e$, and horizontal diamonds for $\bar \nu_e$.
The solid line histograms in 1D figures show the $\nu$ fluxes,
and the dotted ones the $\bar\nu$ fluxes for 1D-multipole calculation.
The solid lines in 3D-nomag figures show the $\nu$ fluxes, 
and the dotted lines the $\bar\nu$ fluxes from Ref.~[13].
The neutrino fluxes are integrated in the energy range of 0.1~--~0.3~GeV,
0.3~--~1~GeV, and 1~--~3.1~GeV, and averaged over all the
azimuth angles.
${\rm Cos}({\rm zenith})=1$ is for the downward going neutrinos.
The results of 1D calculation is also plotted in the 3D figure
as the solid line histogram for $\nu$ and dotted line histogram for
$\bar\nu$ for the comparison.
}
\label{zdep-a}
\end{figure}

\begin{figure}[tbh]
\centerline{\epsfxsize=17cm\epsfbox{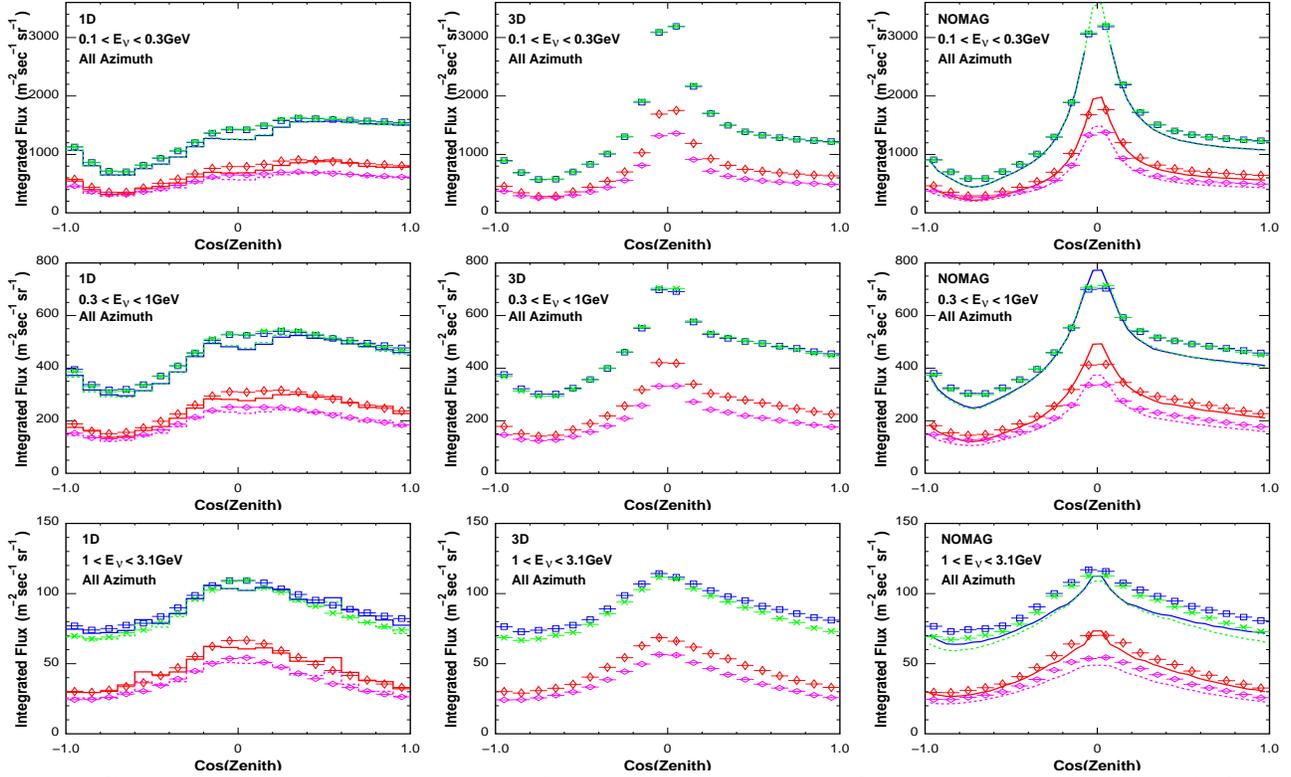}}
\caption{Zenith angle dependence of atmospheric neutrino flux
calculated in the 1D(left), 3D(center), and 3D-nomag(right) schemes for HML.
Squares are for $\nu_\mu$, asterisks for $\bar \nu_\mu$, 
vertical diamonds 
for $\nu_e$, and horizontal diamonds for $\bar \nu_e$ for 1D,
3D, and 3D-nomag calculations.
The solid line histograms in 1D figures show the $\nu$ fluxes, 
and the dotted ones the $\bar\nu$ fluxes for 1D-multipole calculation.
The solid lines in 3D-nomag figures show the $\nu$ fluxes, 
and the dotted lines the $\bar\nu$ fluxes from Ref.~[13].
The neutrino fluxes are integrated in the energy range of 0.1~--~0.3~GeV,
0.3~--~1~GeV, and 1~--~3.1~GeV, and averaged over all the
azimuth angles.
${\rm Cos}({\rm zenith})=1$ is for the downward going neutrinos.
The results of 1D calculation is also plotted in the 3D figure
as the solid line histogram for $\nu$ and dotted line histogram for
$\bar\nu$ for the comparison.
}
\label{sno-zdep-a}
\end{figure}

In the 1-dimensional approximation, and
without the geomagnetic cutoff,
we expect larger atmospheric neutrino flux for 
the horizontal direction than for the vertical direction.
Since the average first interaction point of cosmic rays
is $\sim$ 100 ${\rm g/cm^2}$ in column density, 
the inclined cosmic rays produce pions in a higher altitude 
than the vertical ones.
When $\pi$--$\mu$ decays take place in the dense air, 
the resulting neutrino flux is reduced, since 
the interaction probability of pions with the other air nuclei
and the energy loss of the muons increases with the air density.
Note, however, the muon energy loss is more important 
than the interaction probability of pions 
in the energy region in which we are interested ($\lesssim$~1~GeV).

The geomagnetic cutoff modifies the zenith angle dependence of 
the atmospheric neutrino flux at low energies ($\lesssim$~3~GeV).
In MML, even high rigidity particles ($\gtrsim$~35~GV)
do not pass the geomagnetic cutoff test 
in the near-horizontal easterly directions, while
relatively low rigidity particles ($\sim$ 11~GV) pass the test
in the near-vertical directions.
Thus the atmospheric neutrino flux for horizontal directions 
is lower than that for the neighboring directions, even 
after averaging over all azimuth angles.
In HML, we expect the downward going neutrino flux is larger than 
the upward going neutrino flux, since averaged 
cutoff rigidity is lower for down going directions.

We show the zenith angle variation calculated in the 1D, 3D, and 
3D-nomag schemes in Figs.~\ref{zdep-a} and ~\ref{sno-zdep-a}
for MML and HML respectively to compare the zenith angle dependence 
of the atmospheric neutrino flux among different calculation schemes.
The neutrino fluxes are integrated in the energy range of 0.1~--~0.3~GeV,
0.3~--~1~GeV, and 1~--~3.1~GeV, and averaged over all 
azimuth angles.

The zenith angle variations of the atmospheric neutrino fluxes calculated 
in the 1D and 1D-multipole schemes are well explained by the incident angle 
of the primary cosmic ray and the geomagnetic cutoff as above.
In the 3D and 3D-nomag calculations, however, 
there is a large enhancement of neutrino fluxes in the 
near-horizontal directions, which is not seen in the
1D and 1D-multipole calculations.
This horizontal enhancement is seen in both MML and HML.
The horizontal enhancement of atmospheric neutrino flux 
was first reported by Battistoni et al.~\cite{battis},
and Lipari gave an explanation in terms of geometry~\cite{lipari-ge}.

%\\\
For the comparison with the results of Battistoni et al.~\cite{battis},
we plotted their results alongside our 3D-nomag results,
since they considered the geomagnetic field outside the atmosphere
to calculate the geomagnetic cutoff 
but they did not apply the geomagnetic field in the atmosphere.
We found the horizontal enhancement is $\sim$~10~\% larger than ours.
However, this is not considered to be the result of the difference 
between the multipole and dipole geomagentic cutoff schemes.
The 1D calculation gives a larger flux than the 1D-multipole 
at near-horizontal directions in our study.
Therefore, we conclude that the larger horizontal 
enhancement in the results Battistoni et al. can be explained by 
the difference of hadronic interaction model from ours,
especially that of transverse momentum of secondary pions.

\subsection{East-West effect}
\label{east-west}

\begin{figure}[tbh]
\centerline{\epsfxsize=17cm\epsfbox{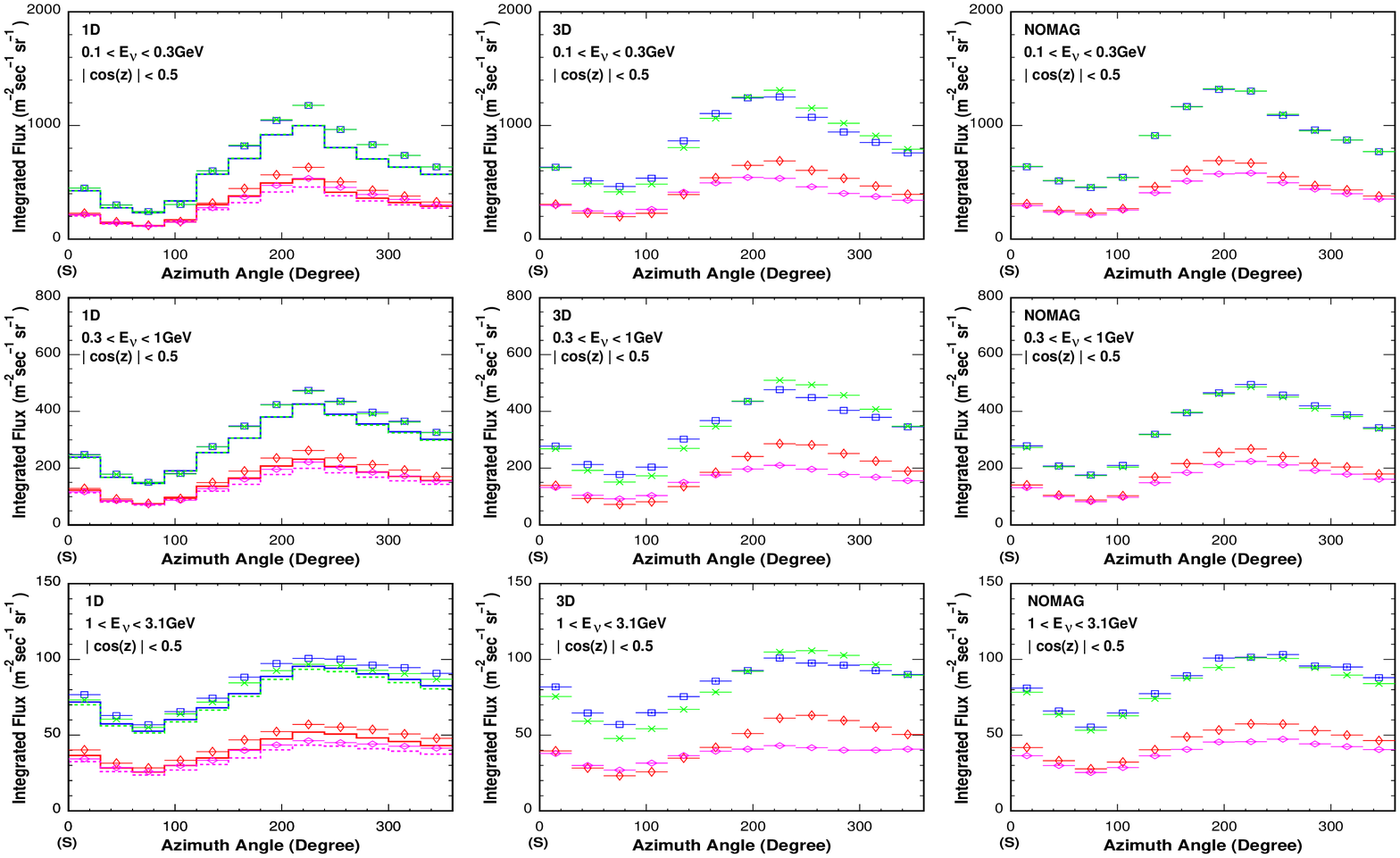}}
\caption{Azimuth dependence of atmospheric neutrino fluxes
calculated in the 1D(left), 3D(center), and 3D-nomag(right) schemes for MML.
Squares are for $\nu_\mu$, asterisks for $\bar \nu_\mu$, vertical diamonds
for $\nu_e$, and horizontal diamonds for $\bar \nu_e$.
The neutrino fluxes are integrated over the energy range of 
0.1~--~0.3~GeV, 0.3~--~1~GeV, and 1~--~3.1~GeV, 
and averaged over the zenith angles:
$|\cos(\theta_{\rm Zenith})|<0.5$.
The results of 1D-multipole are plotted in 1D figures;
the solid lines show the $\nu$ fluxes, and the dotted line 
the $\bar\nu$ fluxes.
${\rm Azimuth}=0, 90, 180, 270,$ 
are the magnetic southerly, easterly, northerly, and westerly directions,
respectively.
}
\label{azim-z0.5}
\end{figure}

\begin{figure}[tbh]
\centerline{\epsfxsize=17cm\epsfbox{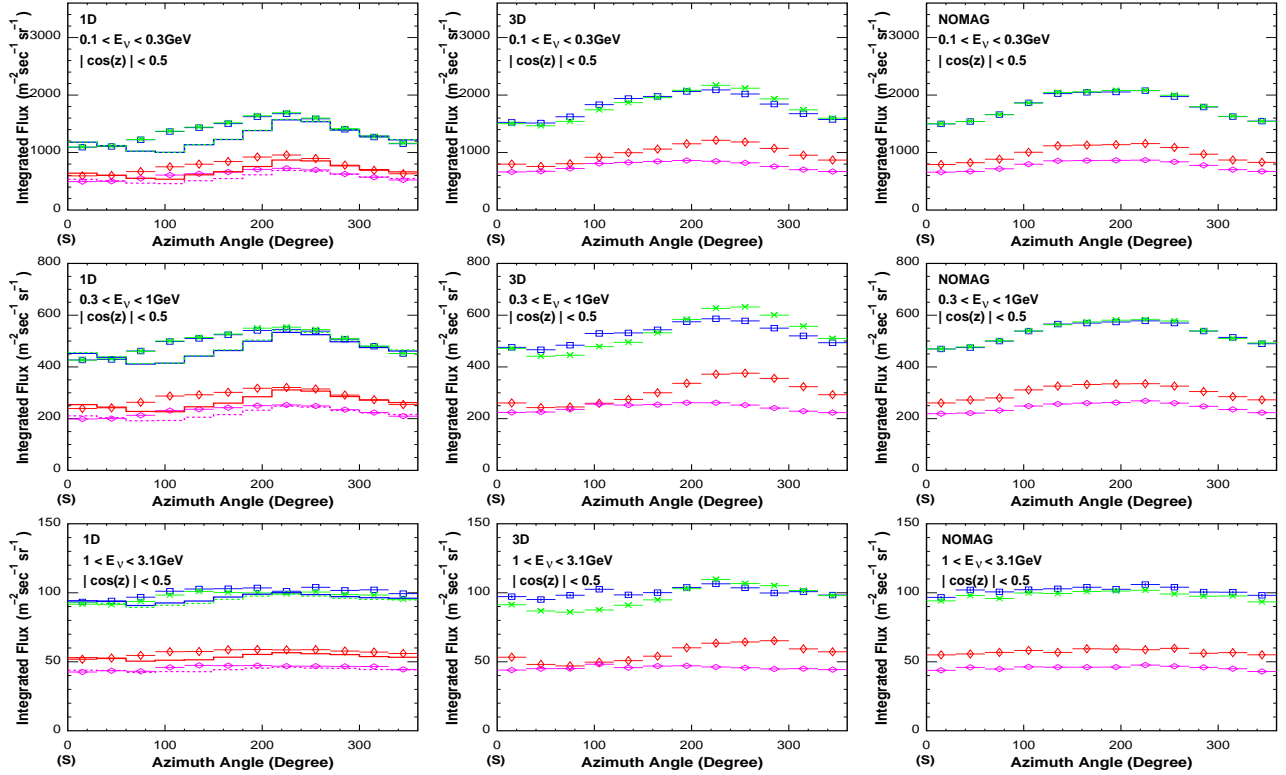}}
\caption{Azimuth dependence of atmospheric neutrino flux
calculated in the 1D(left), 3D(center), and 3D-nomag(right) procedures for HML.
Squares are for $\nu_\mu$, asterisks for $\bar \nu_\mu$, vertical diamonds
for $\nu_e$, and horizontal diamonds for $\bar \nu_e$.
The neutrino flux is integrated over the energy range of 0.1~--~0.3~GeV,
0.3~--~1~GeV, and 1~--~3.1~GeV, and averaged over the zenith angles:
$|\cos(\theta_{\rm Zenith})|<0.5$.
The results of 1D-multipole are plotted in 1D figures;
the solid lines show the $\nu$ fluxes, and the dotted line 
the $\bar\nu$ fluxes.
${\rm Azimuth}=0, 90, 180, 270,$ 
are the magnetic southerly, easterly, northerly, and westerly directions
respectively.
}
\label{sno-azim-z0.5}
\end{figure}

We show the azimuth variation of the neutrino fluxes
calculated in 1D, 3D, and 3D-nomag schemes  
in Fig.~\ref{azim-z0.5} and Fig.~\ref{sno-azim-z0.5} 
for MML and HML, respectively.
The azimuth variation calculated in 1D-multipole scheme 
is also shown with that in the 1D scheme as a histogram.
The neutrino flux is integrated over the energy range of 0.1~--~0.3~GeV,
0.3~--~1~GeV, and 1~--~3.1~GeV, and averaged over the zenith angles:
$|\cos(\theta_{\rm Zenith})|<0.5$.

The azimuthal variation of the neutrino fluxes is 
determined only by the geomagnetic cutoff in the 1-dimensional
approximation, 
and we expect only a small deviation from that in the 3-dimensional 
schemes. 
(For detailed discussions, see Lipari et al.~\cite{lgs}.)
The difference between 1D, 3D, and 3D-nomag is small since 
they use the same geomagnetic cutoff scheme.
The difference between the dipole and multi-pole 
geomagnetic cutoff scheme is also small.
Note that there is an experimental study of the azimuth variation of
the atmospheric neutrino flux\cite{futagami}, although the statistics
in this study is small.

In both MML and HML,
the atmospheric neutrinos fluxes are larger in the 
westerly directions ($180^\circ < {\rm azimuth} < 360^\circ$) than the 
easterly directions ($0^\circ < {\rm azimuth} < 180^\circ$) due to the 
lower cutoff rigidity for the westerly directions.
However, the difference between the westerly direction and
the easterly direction is smaller in HML than in MML.
Since the corresponding energy to the cutoff rigidity is 
near or even lower than the pion production threshold 
of cosmic rays in HML,
the effect of the geomagnetic cutoff is small.

We would like to note, however, that there is a feature 
which is not explained by the geomagnetic cutoff only.
The azimuth variation of $\bar\nu_\mu$ and $\nu_e$ fluxes  
is larger than that of $\nu_\mu$ and $\bar\nu_e$ fluxes
only in the 3D calculation, and this is not seen 
in other calculation schemes.
The differences betweeen $\nu_\mu$ and $\bar\nu_\mu$,
and $\nu_e$ and $\bar\nu_e$ are considered to result from 
the curvature of muons in the geomagnetic field.
This feature is seen both in HML and MML.

\begin{figure}[tbh]
\centerline{\epsfxsize=17cm\epsfbox{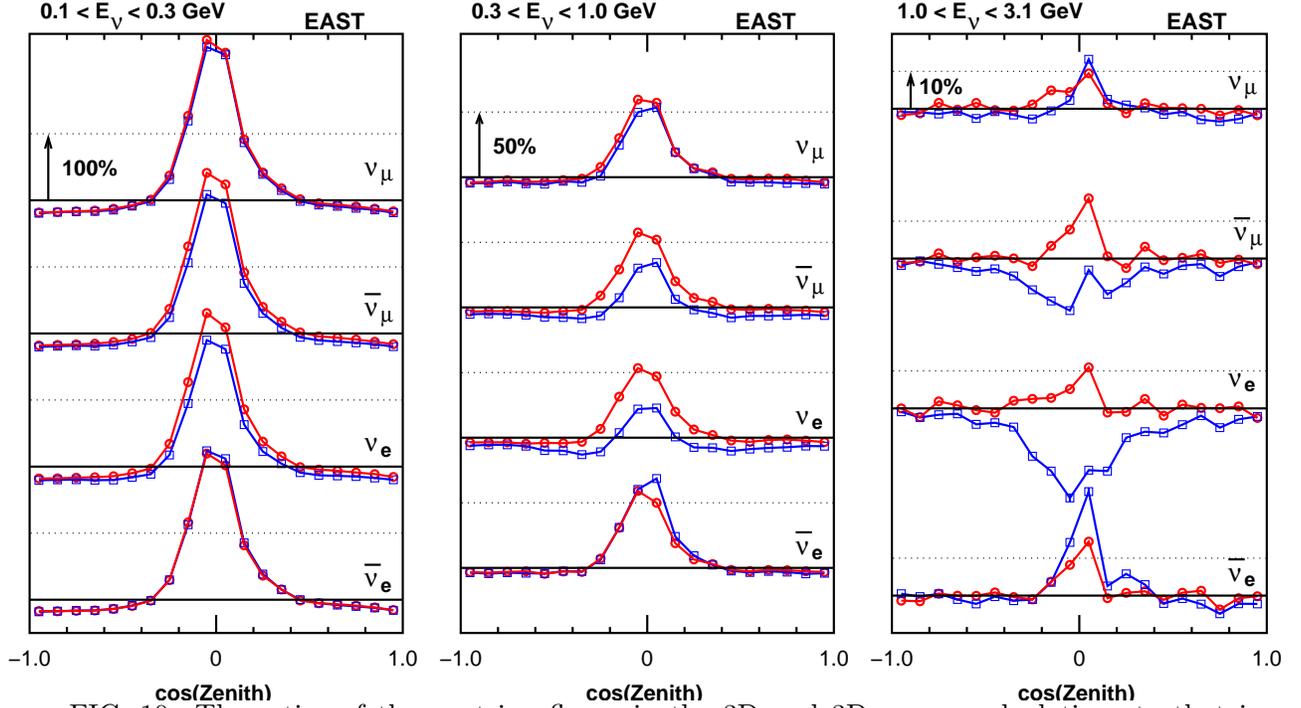}}
\caption{The ratios of the neutrino fluxes in the 3D and 3D-nomag 
calculations to 
that in the 1D calculation for the MML
for 3 energy bands: 0.1~--~0.3~GeV (left), 0.3~--~1~GeV (center), 
and 1~--~3.1~GeV (right)
in the easterly directions, 
Squares indicate 3D to 1D ratios and circles 3D-nomag to 1D ratios.
The scales are different for each energy band.
}
\label{ratio-zdep-e}
\end{figure}
\begin{figure}[tbh]
\centerline{\epsfxsize=17cm\epsfbox{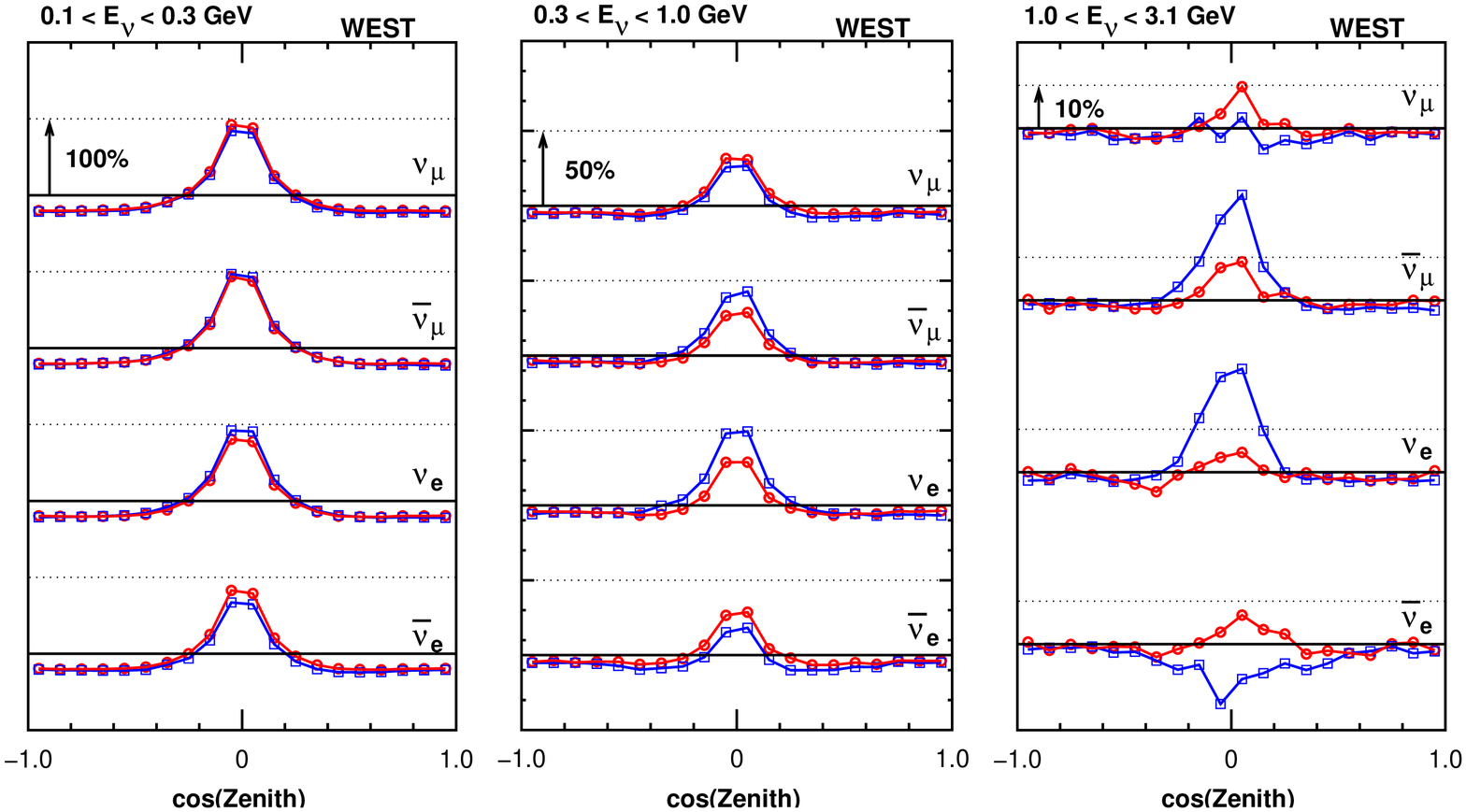}}
\caption{The ratios of the neutrino fluxes in the 3D and 3D-nomag calculations to 
that in the 1D calculations for MML
for 3 energy bands: 0.1~--~0.3~GeV (left), 0.3~--~1~GeV (center), 
and 1~--~3.1~GeV (right)
in the westerly directions.
Squares indicate 3D to 1D ratios and circles 3D-nomag to 1D ratios.
The scales are different for each energy band.
}
\label{ratio-zdep-w}
\end{figure}

In order to study the horizontal enhancement in more detail, 
we have taken the flux ratio between the 3D and 3D-nomag calculations 
and the 1D calculation for MML. 
The flux ratio is shown separately in Fig.~\ref{ratio-zdep-e} 
for easterly directions
(S-E-N 180$^\circ$ in azimuth angle),
and in Fig.~\ref{ratio-zdep-w} for westerly directions 
(N-W-S 180$^\circ$ in azimuth angle).
Note that the unit of the vertical axis is different 
in these ratio-figures. 

Firstly, the amplitude of the horizontal enhancement is different 
in easterly and westerly directions for both 3D and 3D-nomag
calculations.
In the near-horizontal easterly directions,
it is $\sim$ 200~\% for
0.1~--~0.3~GeV, and $\sim$ 50~\% for 0.3~--~1.0~GeV, while
in the westerly directions,
it is $\sim$ 100~\% for 0.1~--~0.3~GeV,
and $\sim$ 30\% for 0.3~--~1~GeV.
The large amplitude of the horizontal enhancement for easterly 
direction is caused by the 
low neutrino flux in 1D calculation in near-horizontal easterly 
directions 
due to the high cutoff rigidity ($\gtrsim$ 35GV).
The `3D effects' work to smear out such a quick variation,
and the ratio of the 3D flux to 1D one is larger for the region
with higher cutoff rigidity if the zenith angle is the same.

In the energy range of 1.0~--~3.1~GeV, however,
the horizontal enhancement becomes small and the 
difference between 3D and 3D-nomag calculation becomes apparent.
In the 3D calculation,
$\nu_\mu$ and $\bar\nu_e$ fluxes are enhanced, while
$\bar\nu_\mu$ and $\nu_e$ fluxes are suppressed for easterly 
horizontal direction, and $\bar\nu_\mu$ and $\nu_e$ fluxes are
enhanced, while  $\nu_\mu$ and $\bar\nu_e$ fluxes are suppressed
for westerly horizontal direction.
These feature is not seen in the 3D-nomag calculation.
Remember the fact that both the $\bar \nu_\mu $ and $\nu_e$ are produced in 
a $\mu^+$ decay, 
while both the $\nu_\mu$ and $\bar\nu_e$ are produced in a $\mu^-$ decay.
These enhancement and suppression are related to the muon curvature
in the geomagentic field.

A muon is produced in a direction following the $P_t$ 
distrubution of pions in the hadronic interaction of the 
parent cosmic ray.
The directions of the potential parent cosmic rays 
of a muon distribut in a axisymmetric distribution around 
the muon direction at the muon production point, ignoring the
bending of pions in the geomagnetic field.
For near-horizontal muons, therfore,
some of the potential parent cosmic rays are shaded by the Earth.
Since a muon changes its direction by $\sim 5^\circ$ 
within the average life time in the geomagnetic field,
the shading of the cosmic ray by the Earth works different 
way depending on the direction and chage of muons.
For a easterly $\mu^+$ and westerly $\mu^-$, 
the shading by the Earth works more effectively, 
and the neutrinos produced by these muons are suppressed.
For easterly $\mu^-$ and westerly $\mu^+$, 
the shading by the Earth works less effectively,
and the neutrino produced by these muons are enhanced.
Note, the shading by the Earth reduce the neutrino flux near 
the horizontal directions, but is not seen in the resulting 
neutrino flux due to the large horizontal enhancement.

There is an additinal effect related to the geomagnetic cutoff.
The geomagnetic cutoff applied to the parent cosmic ray is 
different between 3D and 3D-nomag calculations due to the muon curvature. 
Generally speaking, a higher cutoff rigidity is applied to a $\mu^+$,
and a lower rigidity cutoff is applied to a $\mu^-$ irrespective
of direction.
However, this effect is not evident except for the easterly 
and near-horizontal directions, where the cutoff rigidity rapidly 
increase toward the horizontal direction.

The amplitudes of the enhancement or suppression in 
geomagnetic field are
$\sim$~5~\% for the $\nu_\mu$, and
$\sim$~10~\% for the $\bar\nu_\mu$,
20~$\sim$~30~\% for $\nu_e$ and 
10~$\sim$~20~\% for $\bar\nu_e$,
over a wide energy range in the near-horizontal, 
westerly or easterly directions.
These amplitude could be understood by the muon curvature effect
explained above.
Note Lipari pointed the importance of the magnetic field 
after the geomagnetic cutoff in Ref.~\cite{lipari-ew}.

\begin{figure}[tbh]
\centerline{\epsfxsize=17cm\epsfbox{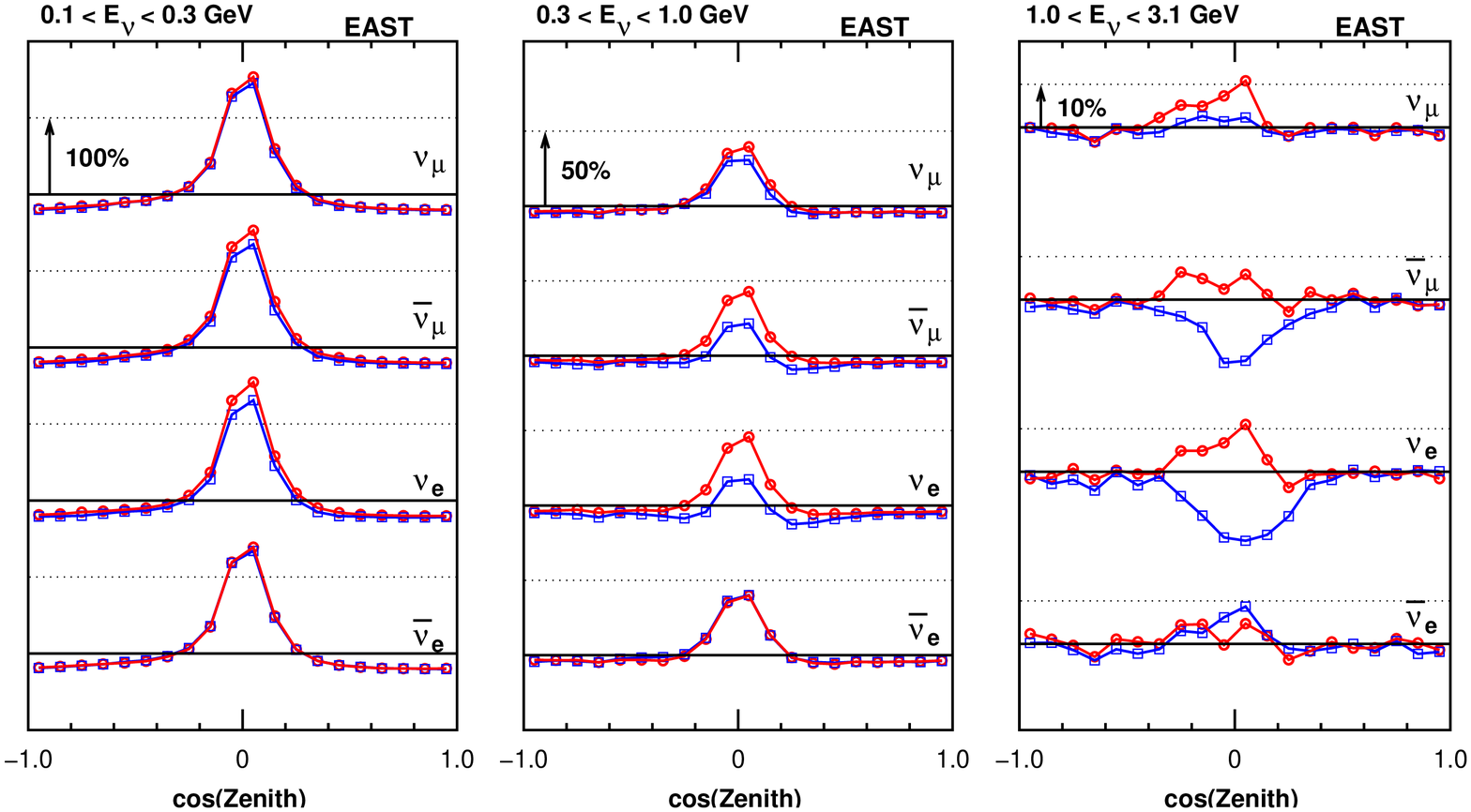}}
\caption{The ratios of the neutrino fluxes in the 3D and 3D-nomag calculations to 
that in the 1D calculation for the HML
for 3 energy bands: 0.1~--~0.3~GeV (left), 0.3~--~1~GeV (center), 
and 1~--~3.1~GeV (right)
 in the easterly directions.
Squares indicate 3D to 1D ratios and circles 3D-nomag to 1D ratios.
The scales are different for each energy band.
}
\label{sno-ratio-zdep-e}
\end{figure}
\begin{figure}[tbh]
\centerline{\epsfxsize=17cm\epsfbox{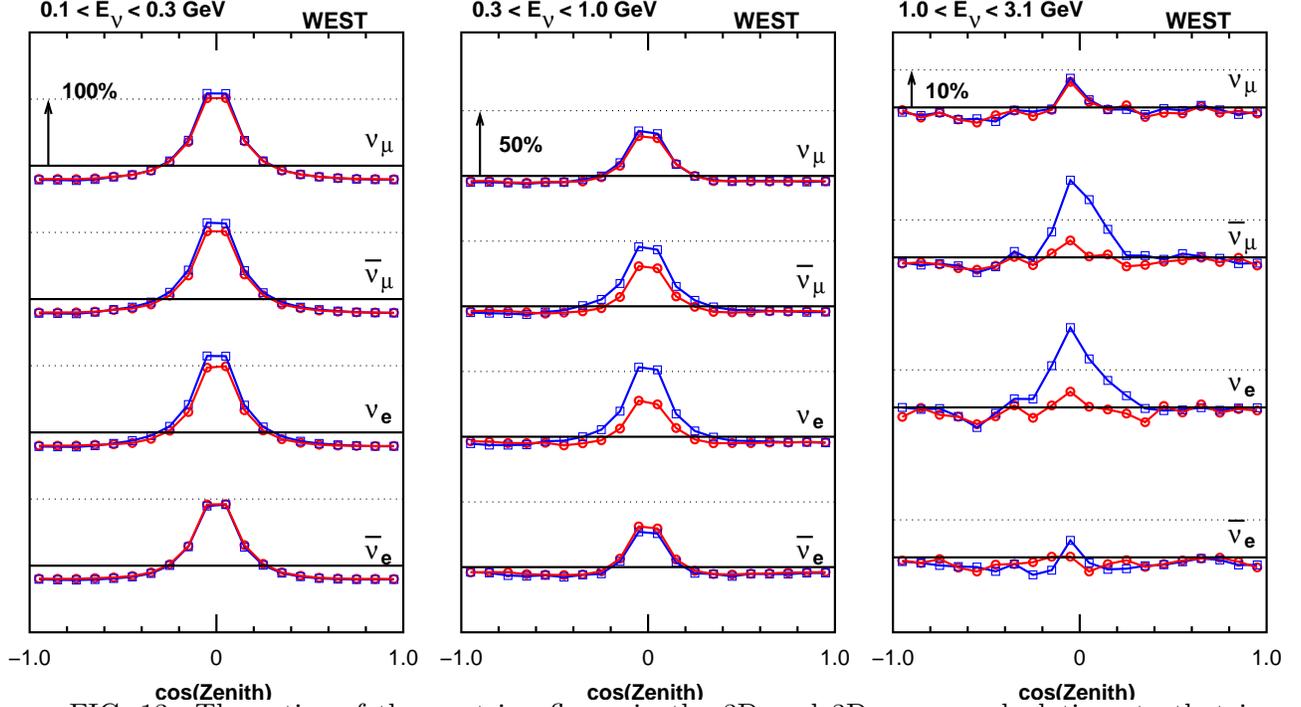}}
\caption{The ratios of the neutrino fluxes in the 3D and 3D-nomag calculations to 
that in the 1D calculation for the HML 
for 3 energy bands: 0.1~--~0.3~GeV (left), 0.3~--~1~GeV (center), 
and 1~--~3.1~GeV (right)
 in the westerly directions.
Squares indicate 3D to 1D ratios and circles 3D-nomag to 1D ratios.
The scales are different for each energy band.
}
\label{sno-ratio-zdep-w}
\end{figure}

For the HML, the flux ratio is also calculated and shown 
separately in Fig.~\ref{sno-ratio-zdep-e} for easterly directions
(S-E-N 180$^\circ$ in azimuth angle),
and Fig.~\ref{sno-ratio-zdep-w} for westerly directions 
(N-W-S 180$^\circ$ in azimuth angle).
Note that the unit of the vertical axis is different 
in these ratio-figures. 

The general feature is the same as in the MML, but
the horizontal enhancement is larger for lower neutrino
energies, 
and the difference between 3D and 3D-nomag calculation is 
seen in near-horizontal directions for $\gtrsim$ 1~GeV.
However, the difference between easterly and westerly directions
is smaller than that in the MML, since the effect of rigidity 
cutoff is small even in the near-horizontal directions.
It is $\sim$ 100~\% for 0.1~--~0.3~GeV,
and $\sim$ 30\% for 0.3~--~1~GeV for both directions.
Also the differences of the fluxes in 3D and 3D-nomag calculation are
similar to those in MML, but a little smaller.

\subsection{Neutrino Production Height}
\label{production-height}

\begin{figure}[tbh]
\centerline{\epsfxsize=17cm\epsfbox{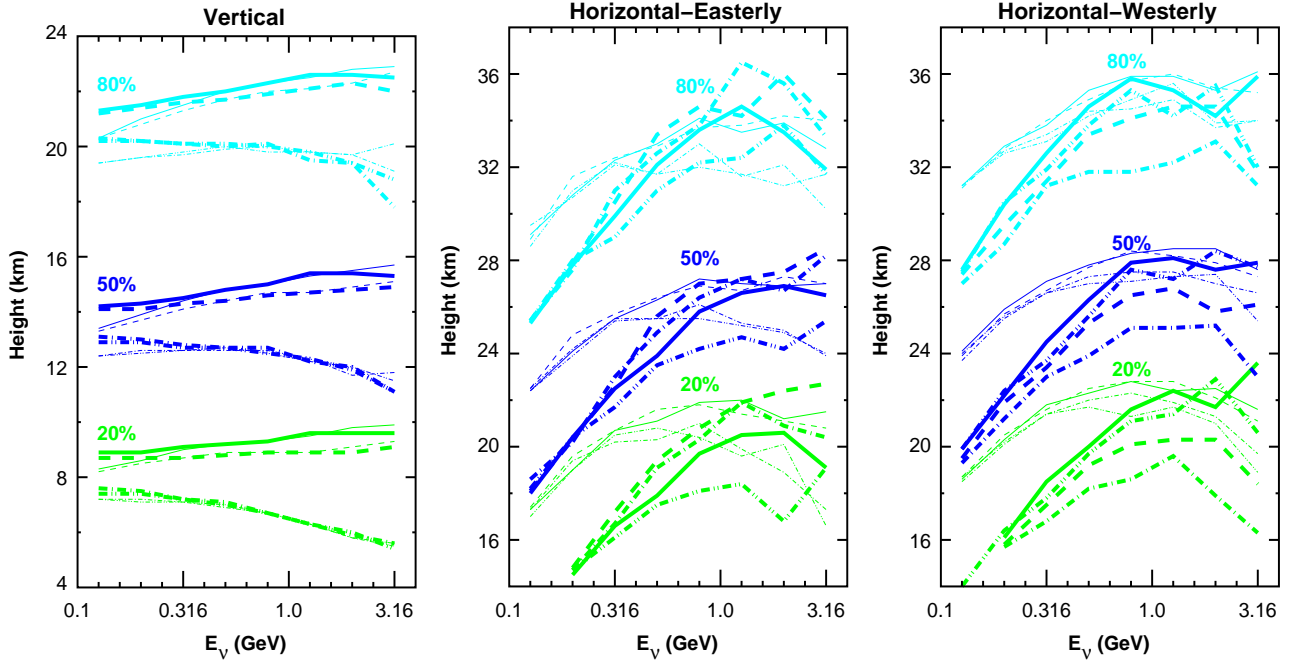}}
\caption{Constant accumulation probability line for neutrino
production height for near-vertical($\cos({\rm zenith}))>0.9$)(left),
near-horizontal ($|\cos({\rm zenith})|<0.1$) easterly (center),
and near-horizontal westerly (right) directions for MML.
Thick solid lines are for $\nu_\mu$, 
thick dashed lines for $\bar\nu_\mu$, 
thick dash-dotted lines for $\nu_e$ , and 
thick dash-double-dotted lines for $\bar\nu_e$ by the 3D calculation.
Thin solid lines are for $\nu_\mu$, 
thin dashed lines for $\bar\nu_\mu$, 
thin dash-dotted lines for $\nu_e$ , and 
thin dash-double-dotted lines for $\bar\nu_e$ by the 1D calculation.
}
\label{height-kam}
\end{figure}

\begin{figure}[tbh]
\centerline{\epsfxsize=17cm\epsfbox{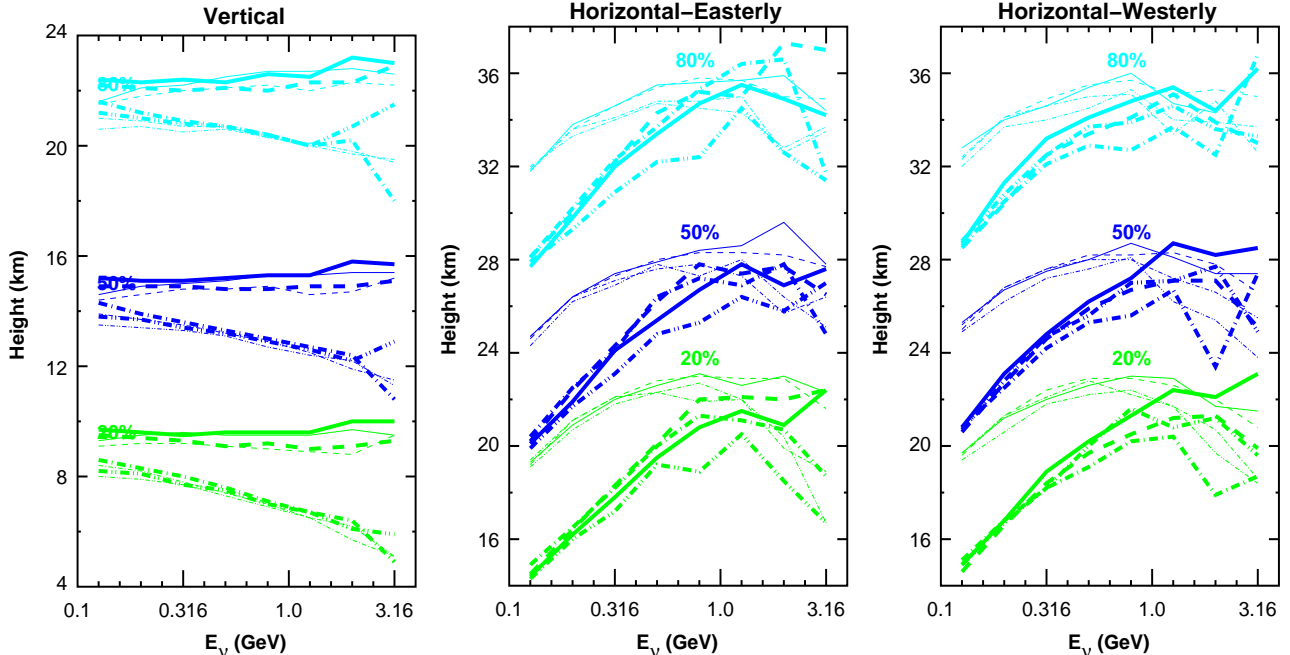}}
\caption{Constant accumulation probability line for neutrino
production height for near-vertical($\cos({\rm zenith}))>0.9$)(left),
near-horizontal ($|\cos({\rm zenith})|<0.1$) easterly (center),
and near-horizontal westerly (right) directions for HML.
Thick solid lines are for $\nu_\mu$, 
thick dashed lines for $\bar\nu_\mu$, 
thick dash-dotted lines for $\nu_e$ , and 
thick dash-double-dotted lines for $\bar\nu_e$ by the 3D calculation.
Thin solid lines are for $\nu_\mu$, 
thin dashed lines for $\bar\nu_\mu$, 
thin dash-dotted lines for $\nu_e$ , and 
thin dash-double-dotted lines for $\bar\nu_e$ by the 1D calculation.
}
\label{height-sno}
\end{figure}

As we have already discussed in section \ref{sec-zdep},
the production height of atmospheric neutrinos is mainly
determined by the zenith angle of incoming cosmic rays
in the 1D calculation.
The cutoff rigidity also gives an additional effect to the production
height for low energy neutrinos.
The production height of a fixed energy neutrino is lower for higher 
energy parent cosmic rays,
since the interaction-decay cascade extends deeper into the 
atmosphere when it is initiated by higher energy cosmic rays.
The production height is also different for different kinds of neutrino.
It is lower for $\nu_e$ and $\bar\nu_e$ than for 
$\nu_\mu$ and $\bar\nu_\mu$,
because the former are mainly produced only in the decay of muons and 
the latter are produced both in the muon and pion decays,
and the muons are mainly produced in the pion decay.

In order to study the difference of the production height
between the 1D and 3D calculations,
we integrate it from ground level to the top of the atmosphere 
for the 1D and 3D calculations.
We show the accumulated probabilities of 20~\%, 
50~\%(median), and 80~\% in Figs.~\ref{height-kam} 
and \ref{height-sno}.

We find that the production heights in the 3D and 1D calculations
are almost identical for the near-vertical directions.
Also they roughly agree each other for the near-horizontal directions
at high energies ($\gtrsim$~1~GeV).
However, 
the production height in the 3D calculation is apparently lower than 
that in the 1D calculation for near-horizontal direction and
neutrino energies $<$~1~GeV.
Despite of this difference, the essential discussions for 
the production height in the 1D calculation
can be applied to the 3D calculation.
The production height for near-vertical direction is lower than 
that for near-horizontal direction.
Also the effect of geomagnetic cutoff on the production height 
can be seen in the comparison of MML
and HML, and in the comparison for easterly and 
westerly near-horizontal directions.

From these figures, we find that the neutrino production heights 
calculated in the 1D and 3D calculations agree with each other 
for $>$ 0.3GeV 
in near-vertical direction, and for $>$ 1GeV in near-horizontal 
directions.
Note that at high energies ($\gtrsim$ 2 GeV), our calculation also
suffers from small statistics.
However, in the near-horizontal direction and for $<$~1~GeV,
the production height is lower in the 3D calculation than that in 
the 1D calculation. 
The difference of the production height between the decay products of $\mu^-$ 
($\nu_\mu$ and $\bar\nu_e$) and those of $\mu^+$ 
($\bar\nu_\mu$ and $\nu_e$) is also seen in the 3D calculation in the
energies of $\gtrsim$~0.3~GeV.
The production height of $\nu_\mu$ or $\bar\nu_e$ is 
higher than that of $\bar\nu_\mu$ or $\nu_e$ for westerly direction,
and the production height of $\nu_\mu$ or $\bar\nu_e$ is 
lower than that of $\bar\nu_\mu$ or $\nu_e$ for easterly direction.
This can be understood by the curvature of muons.

\begin{figure}[tbh]
\centerline{\epsfxsize=12cm\epsfbox{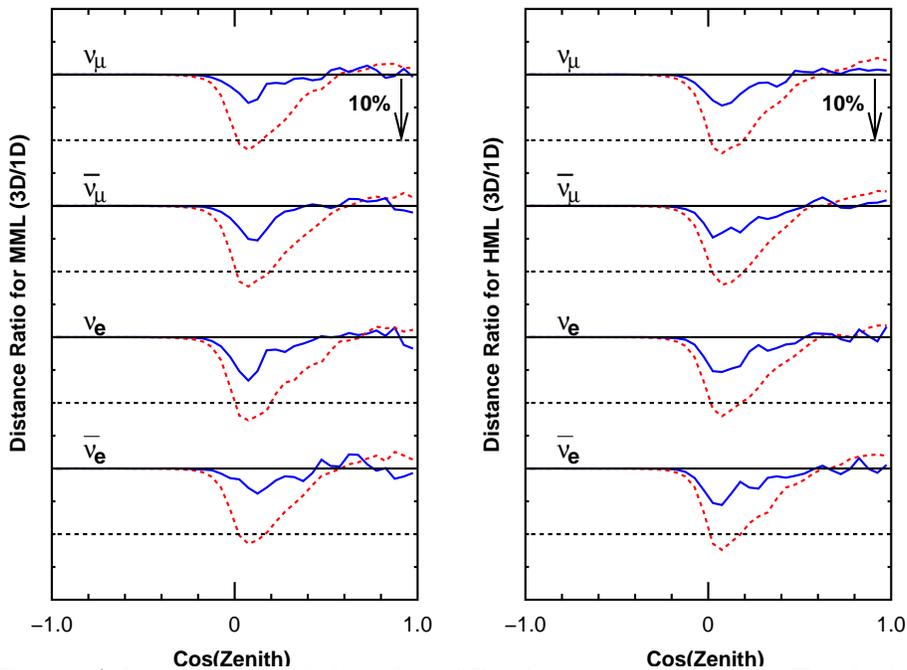}}
\caption{Ratio of the median path length in 3D calculation
to that in 1D calculation both for MML(left) and HML(right).
Solid line show the ratio for 1~GeV neutrinos, and dotted line for
0.3~GeV neutrinos.
$\cos\theta = 1$ denotes the downword direction for neutrinos.
}
\label{distance-ratio}
\end{figure}

Gaisser and Stanev stressed the importance of the production height
and the path length in the study of neutrino oscillations\cite{gs}.
We compare the path length calculated in the 1D and 3D calculations,
converting the median production height to the path length 
by a simple relation:
\begin{equation}
d = \sqrt{(h^2+2R_e h) + (R_e \cos\theta)^2} - R_e \cos\theta~~,
\end{equation}
where $h$ is the height, $R_e$ is the radius of the
Earth, and $d$ is the path length.
We show the ratio of the two path lengths (3D$/$1D) as a function
of cos(Zenith) in Fig.~\ref{distance-ratio}.
In this comparison, we integrated over all the azimuth angles.

At near horizontal directions the production distance of 0.3~GeV 
neutrino is $\sim$~10~\% smaller
for 3D calculation than for 1D at near horizontal directions.
However, the difference is small $\lesssim$ 5~\% for 1~GeV 
neutrinos, as is expected from the comparison of production 
height. 
There is almost no difference between MML and HML in this comparison.

\section{Summary and Discussions}
\label{summary}

We have calculated the flux of atmospheric neutrinos in a 
3-dimensional scheme 
(3D), with the geomagnetic field simplified by a dipole approximation.
We have made two other calculations using the same geomagnetic cutoff scheme: 
one is a 1-dimensional scheme 
(1D) and the other a 3-dimensional calculation without geomagnetic field
in the air (3D-nomag).
Adding to the above three, we have prepared another 1-dimensional 
calculation using the geomagnetic cutoff scheme due to a multi-pole 
expanded geomagnetic field similar to HKKM calculation~\cite{hkkm}
(1D-multipole).

The most remarkable fact is the large enhancement of the low energy 
neutrino flux at near-horizontal directions found both in the 
3D and 3D-nomag calculations in both mid and high magnetic 
latitudes.
This enhancement is already reported by other 3-dimensional 
calculations\cite{battis}\cite{lipari-ge},
and Lipari showed that the enhancement can be explained by 
the geometry\cite{lipari-ge}.

We introduce an explanation which is a little different to that 
by Lipari.
We simplify the 1D-calculation;
assuming that the atmospheric neutrino is produced at a fixed 
height $h$, or on a shpere with the radius of $R_e +h$.
We also ignore the geomagnetic cutoff and the zenith angle 
dependence of production.
As primary cosmic rays arrive uniformly on the sphere,
the neutrino is also produced uniformly at the sphere 
for the downward direction.
The directional distribution is proportional to
\begin{equation}
\cos\theta d\cos\theta dS~~~(\cos\theta > 0),
\label{1dflux}
\end{equation}
where $\theta$ is the zenith angle of the neutrino.
Note that we take 
$\cos\theta > 0$ as the downward direction,
and we integrate over the azimuth angles.

In the 3D calculation,
the neutrino is produced in a little different direction to the 
primary cosmic ray direction.
The directional distribution is calculated by a  
convolution at the production place, and proportional to
\begin{equation}
\int_{\cos\theta' > 0} D(\theta,\theta') \cos\theta' d\cos\theta' d\cos\theta dS .
\label{3dflux}
\end{equation}
where $\theta$ and $\theta'$ are the zenith angle for the neutrino
in case of 3D and 1D calculations respectively,
and $D(\theta,\theta')$ is a dispersion function due to the `3D effects'.
It is important that equation \ref{3dflux} gives a non-zero value at 
$\cos\theta = 0$, unless $D(\theta,\theta')$ is a $\delta$-function.
The ratio of the two expressions,
$
\int_{\cos\theta' > 0} D(\theta,\theta') d\cos\theta' / \cos\theta 
$
has a divergence at $\cos\theta = 0$.
The zenith angle $\theta$ is almost the same as the arrival
direction at the ground, however, 
$\cos\theta$ never be 0 for neutrinos which is observed 
at ground level.
The horizontal direction at the ground level actually corresponds to 
$\cos(\theta) = \sqrt{1-(R_e/(R_e+h))^2}$ at the production sphere.
We do not see a divergence but rather an enhancement of the neutrino flux
at horizontal directions.

For neutrinos with energies $>$~1~GeV,
$D(\theta,\theta')$ is well approximated by the $\delta$-function.
However, as long as $D(\theta,\theta')$ has a extended structure 
more than
$\Delta\theta = 90^\circ - \cos^{-1}(\sqrt{1-(R_E/(R_E+h))^2}) 
\sim 5^\circ$,
a flux enhancement at horizontal directions would be seen.
Thus, although `3D-effects' are small, but they are enhanced by 
the geometry.

When we compare the 1D and 3D calculations averaging over all directions
(Fig.~\ref{alldir-ratios}), 
we found the difference is rather small,
and 3D calculation gives  
$\sim$ 5~\% larger than 1D even at 0.1~GeV
for MML (mid-magnetic-latitudes, SK). 
It is also true that the the difference between the calculation 
with the axisymmetric dipole geomagnetic cutoff (1D) and 
the multipole geomagnetic cutoff (1D-multipole) is also small;
1D gives $\sim$ 5~\% smaller flux than 1D-multipole calculation
Considering these facts, we would be able to conclude that 
the 1-dimensional calculation made in \cite{hkkm} 
is reasonably justified for the MML (mid-magnetic-latitudes, SK)
as for as the average over all directions is concerned.
when the neutrino flux is averaged over all directions.
We note, however, this is not a general statement;
the 3D calculation for the HML (high-magnetic-latitudes, Soudan-II and SNO)
gives $\sim$ 5~\% higher flux at 0.3~GeV and $\sim$ 10~\% higher flux 
at 0.1~GeV.
Thus, the `3d-effects' work more effectively in HML 
than in MML, or for the position with lower cutoff rigidities.

The effect of the geomagnetic field is different for 
the neutrinos produced by $\mu^+$ ($\bar\nu_\mu$ and $\nu_e$) 
and the neutrinos produced by $\mu^-$ ($\nu_\mu$ and $\bar\nu_e$), as 
is also predicted by Lipari\cite{lipari-ew}.
This effect is not as large as the geometric enhancement for $<$ 1~GeV.
However, it gives 5~$\sim$ 30~\% effect depending on the kind of neutrinos
for near-horizontal directions,
and is almost independent of the neutrino energy and magnetic latitude.
Since this is caused by the curvature of muons in the geomagnetic field,
it would affect neutrino fluxes up to energies of $\gtrsim$ 10 GeV.

The comparision of the 3D and 3D-nomag calculations
in the over all direction average is also interesting.
The variation of the cosmic ray shading by the muon curvature 
discussed in section \ref{sec-zdep} works 
different ways in the easterly and westerly directions.
We expect the difference betwen 3D and 3D-nomag calculations
are small due to the compensation of the effect in both directions.
This is true in the calculation in HML; that they agree 
each other within the statistical errors.
In MML, however, the neutrino flux is 2~$\sim$~3~\% smaller in 3D 
calculation than 3D-nomag at $\lesssim$~0.3~GeV even in the 
all direction average (Fig.~\ref{alldir-ratios}).
The coupled effect of muon curvature and geomagnetic cutoff
may explain this fact, since the cutoff rigidity 
($\gtrsim 10$~GV) and the effect works more effectively at MML.

The production heights of the atmospheric neutrino in the 3D
calculation are similar to that in the 1D calculation for $>$~1~GeV.
They are almost identical in the near vertical directions.
In the near-horizontal directions, however, the production height 
in 3D calculation is lower than that in 1D calculation,
and there are apparent differences in the production heights
of between $\nu_\mu$ and $\bar\nu_\mu$,
and between $\nu_e$ and $\bar\nu_e$ due to curvature of muons in 
the geomagnetic field.

The path length of atmospheric neutrino is also compared in the 1D and
3D calculations, integrating all azimuth directions.
The maximum difference is seen at a near horizontal direction,
and is $\sim$~10~\% for 0.3~GeV neutrinos and $\sim$~5~\% neutrinos.
For higher energy neutrinos ($\gg$~1~GeV), we expect very 
small difference between 1D and 3D calculations.

\section{Acknowledgments}
We are grateful to P.~Lipari, A.~Okada, and J.~Nishimura
for useful discussions and comments. 
We thank to S.~Orito and T.~Sanuki for showing us the data before 
publication and for discussions.
We also thank to C.T.~ Taylor for a careful reading of the manuscript.


\begin{references}
\bibitem{fukuda} Kamiokande Collaboration: K.S.~Hirata et al., Phys. Lett. B 280, 146 (1992);\
 Y.~Fukuda et al., Phys. Lett B 335, 237 (1994).
\bibitem{imb} IMB collaboration: D. Casper et al., Phys. Rev. Lett 66, 2561 (1991);\
  R. Bechker-Szendy et al., Phys. Rev. D 46, 3720 (1992).
\bibitem{soudan2} Soudan2 Collaboration: W.W.M.~Allison et al., Phys. Lett. B 446, 1562 (1999). 
\bibitem{macro} MACRO Collaboration: M.~Ambrosio et al., Phys. Lett. B 434, 451 (1998). 
\bibitem{sk} The Super-Kamiokande Collaboration: Y.~Fukuda et al., Phys. Rev. Lett 81, 1562 (1998).
\bibitem{hirata} Kamiokande Collaboration: K.S.~Hirata et al., Phys. Lett B 205, 416 (1988).
\bibitem{gaisser-old}T.K.~Gaisser, T.~Stanev, S.A.~Bludman and H.~Lee, Phys. Rev. Lett. 51, 223 (1983);\
  G.~Barr, T.K.~Gaisser, and T.~Stanev, Phys. Rev. D 39, 3532 (1989).
\bibitem{lpw} J.G.~Learned, S.~Pakvasa, and T.J.~Weiler, Phys. Lett. B 207, 79 (1988).
\bibitem{bw} V.~Berger and K.~Whisnant, Phys. Lett. B 209, 365 (1988).
\bibitem{hhm} K.~Hidaka, M.~Honda and S.~Midorikawa, Phys. Rev. Lett. 61, 1537 (1988).
\bibitem{hkkm} M.~Honda, T.~Kajita, K.~Kahahara, S.~Midorikawa, Phys. Rev. D 54, 4985 (1995),
\bibitem{gaisser-new}V.~Agrawal, T.K.~Gaisser, P.~Lipari, and T.~Stanev, Phys. Rev. D 53, 1314 (1996).
\bibitem{battis} G.~Battistoni et al., Astropart. Phys. 12, 315 (2000).
\bibitem{tserk} Y.~Tserkovnyak et al., hep-ph/9907450
\bibitem{lipari-ge} P.~Lipari, Astropart. Phys. 14, 171 (2000). 
\bibitem{lipari-ew} P.~Lipari, Astropart. Phys. 14, 153 (2000).
\bibitem{kasahara} K.~Kasahara, Proc. of the 24th ICRC, Rome. Vol. 1, 399 (1995). See also the World-Wide-Web address;\\
http://eweb.b6.kanagawa-u.ac.jp/\~kasahara/ResearchHome/cosmosHome/.
\bibitem{webber}W.R.~Webber et. al., 20th ICRC, Moscow vol 1, 325 (1987). %WEBBER79
\bibitem{mass} MASS Collaboration: P.~Pappini, et. al., 23rd ICRC, Calgary 1, 579 (1993). %(MASS)
\bibitem{leap} LEAP Collaboration: W.S.~Seo, et. al., Astrophys.~J., 378, 763 (1987). %LEAP
\bibitem{imax} Imax Collaboration: W.~Menn, et. al., Astrophys.~J., 533, 281 (2000). %(IMAX)%92
\bibitem{caprice} CAPRICE Collaboration: G.~Boezio, et. al., Astrophys.~J., 518, 457, (1999).% (CAPRICE)
\bibitem{bess}BESS Collaboration: T.~Sanuki et al., Astrophys.~J., 545, 1135 (2000).%BESS-proton
\bibitem{ams}AMS Collaboration:~J. Alcaraz et al., c protons, Physics Letters B 490, 27 (2000).%AMS
%H & (He) & CNO in wl-compilation
\bibitem{smith} L.H.~Smith et al., Astrophys.~J. 180 987 (1973).
%H & He in wl-compilation
\bibitem{ryan} M.~Ryan, J.F.~Ormes, V.K.~Balasubrahmanyan, Phys. Rev. Lett. 28 985 (1972).
%P6)
\bibitem{jacee-p}JACEE Collaboration: K.~Asakimori et al., Astrophys.~J., 502, 278 (1998).
\bibitem{moscow} I.P.~Ivanenko et al., Proc. 23rd ICRC, Calgary, 2, 17 (1993).
%P, Fe
\bibitem{aoyama}M.~Ichimura et al., Phys. Rev. D~48, 1949 (1993).
\bibitem{lgs} P.~Lipari, T.~Stanev, and T.K.~Gaisser, Phys. Rev. D58 (1998) 073003).
\bibitem{futagami}The Super-Kamiokande Collaboration: T. Futagami et al., Phys. Rev. Letters 82 (1999) 5194).
\bibitem{gs}T.K.~Gaisser and T.~Stanev, Phys. Rev. D57 (1998) 1977).
\end{references}
\end{document}